\documentclass[pra]{revtex4}

\usepackage{graphicx}
\usepackage{times}
\usepackage{amsmath}

\newcommand{\bfh}{\ensuremath{\boldsymbol{h}}}
\newcommand{\bfR}{\ensuremath{\boldsymbol{R}}}
\newcommand{\fedc}{Fe$^{2+}$}
\newcommand{\zndc}{Zn$^{2+}$}
\newcommand{\kB}{k_{\rm B}}
\newcommand{\kET}{k_{\rm ET}}
\newcommand{\TDA}{T_{\rm DA}}
\newcommand{\subD}{{\rm D}}
\newcommand{\subA}{{\rm A}}
\newcommand{\supi}{{\rm i}}
\newcommand{\supf}{{\rm f}}
\newcommand{\wn}{cm$^{-1}$}

\begin{document}

\title{Electron transfer pathway analysis in bacterial photosynthetic reaction center\footnote{
    in \textit{Chemical Science of $\pi$-Electron Systems}, Eds. T. Akasaka et al., Springer, 2015, Chap. 39.
}}

\author{Hirotaka Kitoh-Nishioka\footnote{
Present address: Department of Chemistry, Graduate School of Science, Nagoya University, Furo-cho, Chikusa-ku, Nagoya 464-8601, Japan. 
\\
E-mail: kito.hirotaka@b.mbox.nagoya-u.ac.jp}
    and Koji Ando\footnote{
E-mail: ando@kuchem.kyoto-u.ac.jp}
}
\affiliation{
Department of Chemistry, Graduate School of Science, Kyoto University, Sakyo-ku, Kyoto 606-8502, Japan. 
}

\begin{abstract}
{A new computational scheme to analyze electron transfer (ET) pathways 
in large biomolecules is presented with applications to ETs
in bacterial photosynthetic reaction center.
It consists of a linear combination of 
fragment molecular orbitals
and an electron tunneling current analysis,
which enables an efficient first-principles analysis of ET pathways
in large biomolecules.
The scheme has been applied to the ET from menaquinone
to ubiquinone via nonheme iron complex in bacterial photosynthetic reaction center.
It has revealed that not only the central \fedc\ ion but also 
particular histidine ligands 
are involved in the ET pathways in such a way to mitigate 
perturbations that can be caused by metal ion substitution and depletion,
which elucidates the experimentally observed insensitivity of the ET rate
to these perturbations.
}
\end{abstract}

\maketitle

\section{Introduction}
\label{sec:intro}

Long-distance electron transfers (ET) play 
essential roles in biological energy conversion \cite{Moser1992,Dutton1994,Winkler1999,GrayWinkler2005,FarverPecht2011}.
The most fundamental are those in photosynthesis.
In photosynthesis, the photon energy is captured
in the form of electronic excitation energy
by light-harvesting antenna
systems composed of aggregates of pigments.
The energy is then funneled to
the \lq special pair\rq, a pair of chromophores, 
in the reaction center embedded in the membrane protein.
From the electronic excited state of the special pair, a series ET processes occurs,
which are followed by transmembrane proton-pumps.
The electrochemical energy thus generated by the gradient of proton concentration 
is utilized at the ATP synthase embedded in the same membrane.
The trinity of long-distance ET, proton-pump, and ATP synthesis is also functioning
in cellular respiration
for energy metabolism.
Thus, the chemical transfers of electrons and protons
are at the core of biological energy conversion.
Both electrons and protons are highly quantum mechanical particles.
Moreover, many degrees of freedom in the proteins are involved.
Therefore,
quantum statistical mechanical description is essential
for microscopic understanding of their transfer processes.
This posts a marked challenge to theoretical and computational chemical physics,
and has been an active area of research along with the rapid advances of
computer technologies in recent years.

There exists a simple but fundamental question in biological ETs
on the roles of protein environments: Are they playing only \lq passive\rq\ roles
of simply holding the
redox moieties at appropriate spatial and orientational arrangements,
or do they play some \lq active\rq\ roles to mediate the ETs by involving their
electronic wave functions as bridge states?
Furthermore, if the protein environments play some active roles,
their details are still unclear;
for instance, 
whether the bridge states contribute mainly to the coherent superexchange mechanism
or to the incoherent step-wise hopping ETs.
This aspect of quantum coherence depends on the subtle microscopic mechanism
involving competition between
the ETs and the nuclear vibrational and conformational relaxations.
Although it remains as an important open question,
it will be out of the scope of this paper.
We rather focus on the aspect of electronic coupling under the assumption of
coherent superexchange mechanism.

In this chapter, we review our recent works on 
long-distance biological ETs with development of a new scheme for theoretical and computational analysis.
In Sec. \ref{sec:2}, we briefly overview the ETs in bacterial photosynthetic reaction center.
After identifying the key quantities in the ET rate formula in Sec. \ref{sec:kET},
we outline in Sec. \ref{subsec:FMO}--\ref{subsec:etpath} the computations of
electronic structure of large molecular systems,
ET matrix elements,
and
ET pathways.
Applications to the ETs in bacterial photosynthetic center are discussed
in Sec. \ref{sec:App}.
Section \ref{sec:concl} concludes.

\section{Electron Transfer in Photosynthetic Reaction Center}
\label{sec:2}

\begin{figure}[b]
\begin{center}
\includegraphics[width=0.45\textwidth]{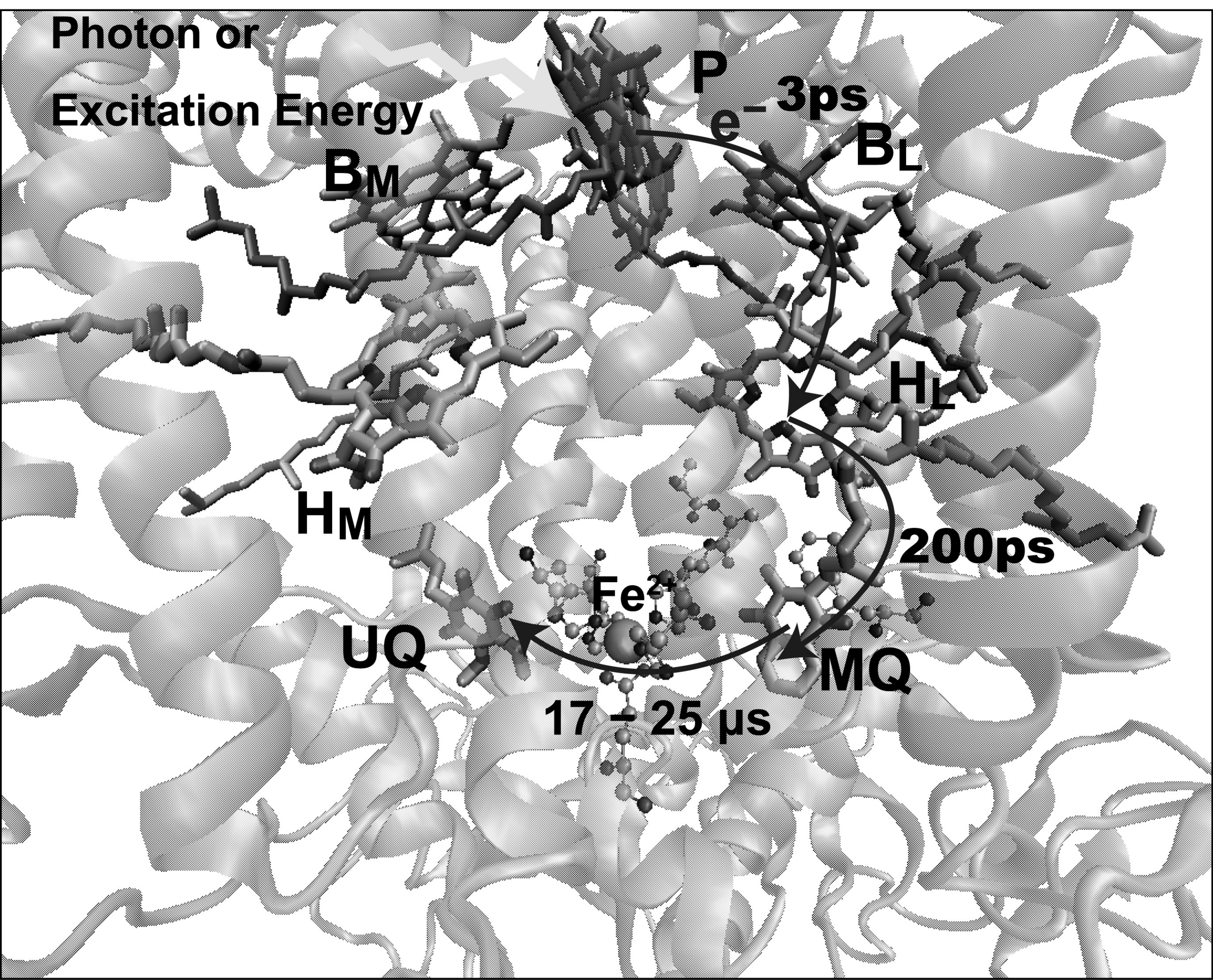}
\caption{Structure of bacterial photosynthetic reaction center.
P is the special pair of bacteriochlorophyll, 
B$_{\rm L}$ and B$_{\rm M}$ are bacteriochlorophylls, 
H$_{\rm L}$ and H$_{\rm M}$ are bacteriopheophytins, 
MQ is menaquinone, and
UQ is ubiquinone.
Between MQ and UQ is a nonheme iron ion complex.
}
\label{fig:1}
\end{center}
\end{figure}

In biological energy conversion processes,
$\pi$-electron systems play essential roles.
For instance, in bacterial photosynthetic reaction center, 
one of the key molecules is bacteriochlorophyll (BChl)
designated by B$_{\rm L}$ and B$_{\rm M}$ in Fig. 1.
(In more details, the structure of BChl is similar to that 
designated by H$_{\rm L}$ in Fig. 2 but with a Mg$^{2+}$ ion
at the center of the tetrapyrrole ring.)
The free-base form indicated by H$_{\rm L}$ in Figs. 1 and 2 
is called bacteriopheophytin (BPhe).
Other key molecules are menaquinone (MQ) and ubiquinone (UQ)
in Fig. 2, and carotenoids.
All these contain
$\pi$-electrons which dominate the
major chemical functions 
including the redox properties.

In the photosynthesis,  
the photons are first captured 
by antenna systems
that consist of aggregates of BChl molecules,
in which the energies are stored
in a form of electronic excitation. 
The excitation energies are then funneled to 
the BChl dimer, called \lq special pair\rq\ (P in Fig. 1),
in the reaction center. 
The electronically excited special pair then ejects
an electron to one of the adjacent BChl molecules (B$_{\rm L}$),
which is followed, coherently or incoherently, by a sequence of ETs 
to BPhe (H$_{\rm L}$)
and to UQ via MQ.
The molecules involved in these last two ET steps are displayed in Fig. 2.
As presented in Fig. 1, these redox centers are embedded in the membrane proteins.
This is thus an intriguing prototype system to examine
the questions on the roles of protein environments
in mediating biological ETs,
as described in Sec. \ref{sec:intro}.
To address these, it is essential to carry out
quantum mechanical analysis at the electronic and atomic level,
for which there exist two-stage basic tasks:
the first is to determine accurate electronic structures of 
large molecules such as membrane proteins,
at least with qualitatively accuracy.
The second is to establish analysis method to clarify the microscopic mechanism
of ETs from the computed electronic wave functions.

\begin{figure}[b]
\includegraphics[width=0.3\textwidth]{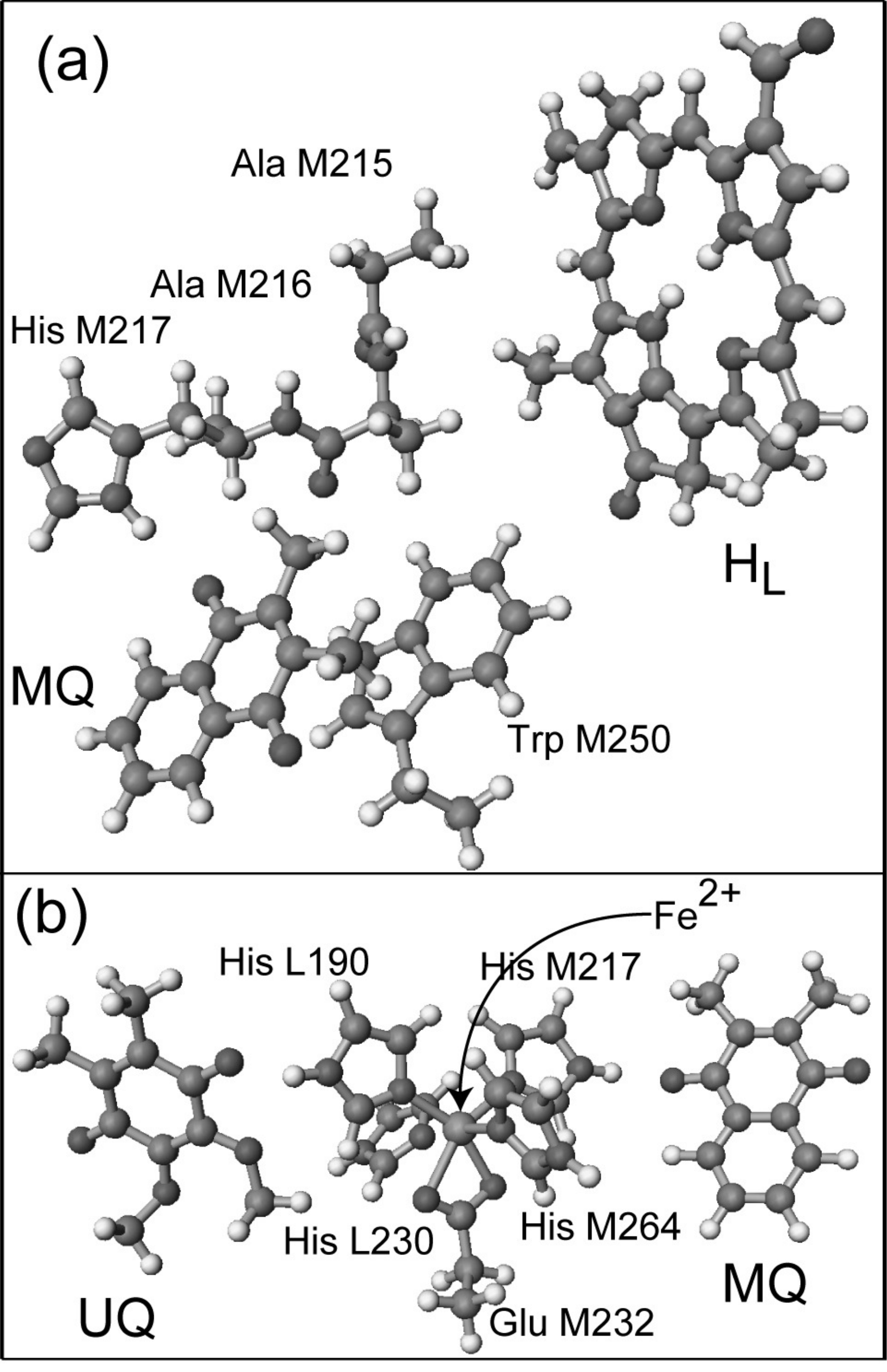}
\caption{Molecular structures (a) around the bacteriopheophytin (H$_{\rm L}$) and menaquinone (MQ)
and (b) the menaquinone and ubiquinone (UQ) via the nonheme iron complex
with four histidine (His) ligands and one glutamate (Glu) ligand.}
\label{fig:2}
\end{figure}

\section{Theory and Computational Methods}
\label{sec:3}

\subsection{Rate of Electron Transfer}
\label{sec:kET}

An essential key to
elucidating ET mechanism is the measurement of rate constant.
The current standard and practically useful theory of ET rate is
the Marcus theory \cite{MarcusSutin1985}, which describes the ET rate $\kET$
with only a few parameters,
the electronic transfer integral $\TDA$,
the reaction free energy $\Delta G^0$, and the nuclear reorganization energy $\lambda$,
\begin{equation}
k_{\rm ET} = \frac{2\pi}{\hbar}
\frac{|\TDA|^2}{\sqrt{4\pi \lambda \kB T}}
\exp \left[-\frac{(\Delta G^0 + \lambda)^2}{4 \lambda \kB T} \right] ,
\label{eq:kET}
\end{equation}
in which $\hbar$ is the Planck constant divided by $2\pi$, 
$\kB$ is the Boltzmann constant, and $T$ is the absolute temperature.
This rate formula was derived as the semiclassical (or high temperature) limit
of thermally averaged Fermi's golden rule which is based on
the time-dependent perturbation theory with respect to the electronic
coupling $\TDA$.
Therefore, application of this formula to long-distance ETs assumes that
the ET occurs as a quantum mechanically coherent process through
the distance of ET.
This aspect still deserves a careful scrutiny \cite{Sumi97}, but in this work
we assume Eq. (\ref{eq:kET}) is appropriate.

Among the three parameters in Eq. (\ref{eq:kET}),
the reaction free energy $\Delta G^0$ can be measured by experiments
most straightforwardly.
The experimental evaluation of the reorganization energy $\lambda$ is
less trivial, but still feasible in various ways.
By contrast, the electronic transfer integral $\TDA$
is the least trivial quantity, which has been estimated in practice only indirectly
from other measurements.
Therefore, theoretical evaluation of $\TDA$
is of particular importance.
Since Eq. (\ref{eq:kET}) was derived with separation of electronic and
nuclear degrees of freedom, that is, with the Condon approximation,
the computation of transfer integral $\TDA$ has been based solely on
the electronic structure calculation under the Born-Oppenheimer approximation.
This adiabaticity aspect again deserves careful examination, but shall be assumed in
this work.
Therefore, the essential problem now is to compute the electronic wave functions
of large molecular systems involved in the biological ETs.

\subsection{Electronic Structure Calculation of Large Molecules}
\label{subsec:FMO}

Although ET reactions can be described basically as a one-electron process,
it is still essential to solve the many-electron problem 
in order to determine 
the major one-electron orbitals.
This is because the most relevant orbitals are 
normally near the highest-occupied molecular orbitals (HOMO).
The computation is highly demanding for large biomolecules, 
but is becoming more and more feasible due to the recent advances of 
hardware power and numerical algorithms.
Two representative methods of large-scale electronic structure calculations
are the fragment molecular orbital (FMO) method \cite{Fedorov2007,Tanaka2014} and 
the divide-and-conquer method \cite{Yang95,Akama2009}.
Here we describe the former which was mainly employed in this work.

\subsubsection{FMO method}

The FMO method \cite{Fedorov2007,Tanaka2014} first decomposes the total system into fragments.
The MOs of each fragment are optimized self-consistently under the Coulomb field of other fragments.
Then, dimer or trimer calculations are carried out 
under the Coulomb field of the optimized monomer fragments
in order to take account of the exchange interactions.
The total energy of the system is computed with
\begin{equation}
E_{\rm total}^{\rm FMO2} = \sum_{I>J} E_{IJ} - (N_{\rm f} - 2) \sum_{I} E_I
\label{eq:fmo2}
\end{equation}
when up to dimer exchange interaction were considered (FMO2), and with
\begin{equation}
E_{\rm total}^{\rm FMO3} = \sum_{I>J>K} E_{IJK} - (N_{\rm f} - 3) \sum_{I>J} E_{IJ} - \frac{1}{2}(N_{\rm f} - 2) (N_{\rm f}-3) \sum_{I} E_I
\label{eq:fmo3}
\end{equation}
when up to trimer exchange interaction were considered (FMO3).
In Eqs. (\ref{eq:fmo2}) and (\ref{eq:fmo3}),
$N_{\rm f}$ is the number of fragments, 
$E_{I}$, $E_{IJ}$, and $E_{IJK}$ denote the energies of fragment monomer, dimer, and trimer,
respectively. 
These formulas 
(\ref{eq:fmo2}) and (\ref{eq:fmo3})
are designed to remove the excess countings of energies
with proper account of self-consistent electronic polarizations.
That is, by carrying out the dimer and trimer calculations under the Coulomb
field of optimized monomer fragments, the excess countings of polarization energies are
also properly removed.
The FMO method gives accurate energies and properties, such as the dipole moments, 
but not the MOs, of
the total system.
However, as will be discussed in Sec. \ref{sec:TDA},
what we need in order to compute the transfer integral $\TDA$ are
the electronic Hamiltonian matrix and the MOs of the total system.
This requirement is fulfilled by the method of linear combinations of FMOs,
the FMO-LCMO method \cite{Tsuneyuki2009,Kobori2013}, which shall be outlined next.

\subsubsection{FMO-LCMO method}
\label{subsec:fmolcmo}

In the FMO-LCMO method,
the \lq intra-fragment\rq\ parts of the Hamiltonian matrix elements 
are computed from
the results of FMO calculations in a form 
similar to Eqs (\ref{eq:fmo2}) and (\ref{eq:fmo3}),
whereas the \lq inter-fragment\rq\ parts are constructed with 
the matrices from 
the dimer or trimer calculations projected to the monomer FMO space.
For instance, in the FMO2 version of the method,
the intra-fragment Hamiltonian matrix elements 
are computed with a formula analogous to Eq. (\ref{eq:fmo2}),
\begin{equation}
H^{\rm (total)}_{Ip,Iq} = \sum_{J \ne I}
\langle \varphi_p^I | \bfh_{}^{IJ} | \varphi_q^I \rangle
- (N-2) 
\langle \varphi_p^I | \bfh_{}^{I} | \varphi_q^I \rangle ,
\label{eq:HtotIpIq}
\end{equation}
and the inter-fragment matrix elements are defined as
\begin{equation}
H^{\rm (total)}_{Ip,Jq} = 
\langle \varphi_p^I | \bfh_{}^{IJ} | \varphi_q^J \rangle
\hspace*{1.5em}
(I \ne J), 
\label{eq:HtotIpJq}
\end{equation}
in which
$\varphi_{p}^{I}$
is the $p$-th orbital of fragment $I$.
The Fock (or Kohn-Sham) matrices of fragment $I$ and fragment dimer $IJ$
are denoted by
$\bfh_{}^{I}$ and
$\bfh_{}^{IJ}$, respectively. 
Thus, 
the notation $\langle \varphi_p^I | \bfh_{}^{IJ} | \varphi_q^J \rangle$
represents the dimer Fock (or Kohn-Sham) matrix projected to 
the monomer FMO space.

Because the FMOs of each monomer fragment 
are optimized independently from other fragments,
the FMOs of different fragments are generally not orthogonal.
This non-orthogonality should be taken into account at the diagonalization.
This has been demonstrated to give
accurate approximations to the canonical MOs and their energies of 
the total system \cite{Tsuneyuki2009,Kobori2013,NishiokaAndo2011B,Kitoh-Nishioka2012}.

By deploying the total Hamiltonian matrix and the MOs of large systems 
thus obtained from the FMO-LCMO method,
we have developed a scheme to analyze the long-distance ET pathways
with the bridge Green function method and the tunneling current method.

\subsection{Electronic Coupling Matrix Elements}
\label{sec:TDA}

\subsubsection{Two-state picture in non-adiabatic regime}

Because the electron transfer integral $\TDA$ is approximately proportional
to the overlap between the donor and acceptor orbitals,
it decays rapidly along their distance.
Thus, for long-distance biological ETs, the transfer integral $\TDA$ is
normally small 
such that the electronically non-adiabatic regime is appropriate.
The ET rate of Eq. (\ref{eq:kET}) assumes this non-adiabatic limit
and is based on the time-dependent first-order perturbation theory
in which the amplitude of the acceptor state $C_\subA (t)$ is proportional
to the perturbation $\TDA$.
Therefore, 
the rate of ET,
the raise of
the population $|C_\subA (t)|^2$,
is proportional to 
$|\TDA|^2$.
Another assumption behind Eq. (\ref{eq:kET}) is that the ET is considered
as an effective two-state problem.
This aspect will be considered in Sec. \ref{subsec:BGF}.
After the reduction to the effective two-state problem, 
the static adiabatic energies, $E_1$ and $E_2$,
are obtained by solving the $2 \times 2$ secular equation
with the diagonal matrix elements $H_{\subD}$ and $H_{\subA}$
and the off-diagonal element $\TDA$,
\begin{equation}
E_{2,1} = \frac{H_{\subD} + H_{\subA}}{2} \pm
\frac{1}{2} \sqrt{ \left( H_{\subD} - H_{\subA} \right)^2 + 4 \TDA^2 } .
\label{eq:E12}
\end{equation}
Here we have omitted the dependence on 
the nuclear coordinates $\bfR$ for simplicity:
the electronic Hamiltonian matrix elements,
$H_{\subD} (\bfR)$, $H_{\subA} (\bfR)$, and $\TDA (\bfR)$, 
and hence the adiabatic energies $E_{1,2} (\bfR)$,
all depend on $R$.
Thus, the adiabatic energy splitting defined by
\begin{equation}
\Delta \varepsilon_{12} (\bfR) \equiv E_2 (\bfR) - E_1 (\bfR)
= \sqrt{ \left( H_{\subD}(\bfR) - H_{\subA}(\bfR) \right)^2 + 4 \TDA(\bfR)^2 }
\label{eq:DeltaE12}
\end{equation}
is twice the
transfer integral $\TDA$ 
at nuclear configurations $\bfR_c$ of the diabatic surface crossing
that gives
$H_{\subD}(\bfR_c) = H_{\subA}(\bfR_c)$,
\begin{equation}
\TDA = \Delta \varepsilon_{12} (\bfR_c) / 2 .
\end{equation}
Nevertheless, it is not a trivial task to find the diabatic surface
crossing configurations $\bfR_c$, especially for large proteins
that involve many degrees of freedom.
This is the reason why we consider the generalized Mulliken-Hush (GMH) analysis
and the bridge Green function method that are described next.

\subsubsection{Generalized Mulliken-Hush analysis}

The GMH method \cite{CaveNewton1997} scales the energy splitting
$\Delta \varepsilon_{12}$ at nuclear configurations \textit{off} 
the surface crossing
by a formula
\begin{equation}
\TDA = 
\frac{|\mu_{12}|\; \Delta \varepsilon_{12}}{\sqrt{(\mu_1 - \mu_2)^2 + 4|\mu_{12}|^2}} ,
\label{eq:GMH}
\end{equation}
in which $\mu_1$ and $\mu_2$ are the dipole moments of
the adiabatic states with $E_1$ and $E_2$,
and $\mu_{12}$ is the off-diagonal element.
Thus, the quantities in the right-hand-side are obtained straightforwardly 
from the standard electronic structure
calculations at any, normally the equilibrium, 
nuclear configurations.
The idea behind this formula is an assumption that 
the Hamiltonian matrix elements
and dipole matrix elements scale similarly
for states involved in ETs,
such that the former elements in Eq. (\ref{eq:DeltaE12})
are replaced by the latter elements to assume the scaling factor.
Despite its simplicity, the GMH formula (\ref{eq:GMH}) has been successfully
applied to a number of ET reactions.
In applications to large systems where the computational cost for
the electronic excited state shall be the bottleneck,
the energy splitting $\Delta \varepsilon_{12}$ is replaced by that
of the donor and acceptor MOs from the ground-state calculation.
Accordingly, the dipole matrix elements 
of the donor-acceptor MOs are also applied.
This provides a computationally feasible and reasonably
accurate method for the transfer integrals of large ET systems.

\subsubsection{Bridge Green function method}
\label{subsec:BGF}

For long-distance ETs in biomolecules,
the electronic Hamiltonian matrix can be very large 
involving the intervening molecular parts between
the donor and acceptor sites.
In the bridge Green function (BGF) method \cite{NishiokaAndo2011B}, the electronic Hamiltonian matrix 
is projected
onto the space of a two-level system consisting of the donor and acceptor states.
The reduced information in the remaining part of the system,
the molecular parts that bridge and mediate the ET,
is taken into account via the BGF matrix.
This will be formulated below
in a generalized form of the effective Hamiltonian method.

We first divide the total Hamiltonian matrix to submatrices
of a target space (P-space),
the remaining space (Q-space),
and their off-diagonal space.
The eigenvalue problem with non-orthogonal basis is thus described by
\begin{equation}
\begin{bmatrix}
H_{PP} - E S_{PP} & H_{PQ} - E S_{PQ} \\
H_{QP} - E S_{QP} & H_{QQ} - E S_{QQ} 
\end{bmatrix}
\begin{bmatrix}
c_P \\ c_Q
\end{bmatrix}
=
\begin{bmatrix}
0 \\ 0
\end{bmatrix} ,
\label{eq:HPQ}
\end{equation}
in which $S$ denotes the overlap matrix.
The non-orthogonal formulation is essential because
the FMOs of different fragments are generally not orthogonal,
as noted in Sec. \ref{subsec:fmolcmo}.
Inserting the formal solution for $c_Q$ from the second line of Eq. (\ref{eq:HPQ})
into the first line,
the problem is reduced the projected smaller P-space
with the effective Hamiltonian 
\begin{equation}
H_{\rm eff} = 
(H_{PP} - E S_{PP}) + 
(H_{PQ} - E S_{PQ})
G(E) 
(H_{QP} - E S_{QP}) ,
\label{eq:Heff}
\end{equation}
in which $G(E)$ is the Green function matrix representing the
contribution from the Q-space (bridge-space),
\begin{equation}
G(E) \equiv
(E S_{QQ} - H_{QQ})^{-1} .
\end{equation}
With the definition of the P-space to be the donor-acceptor states or orbitals, 
($\phi_{\subD}, \phi_{\subA}$),
the off-diagonal element of $H_{\rm eff}$ corresponds to the transfer integral $\TDA$,
\begin{equation}
\TDA = 
\sum_I^N
\sum_J^N
\sum_{I_p} {'}
\sum_{J_q} {'}
( H_{{\subD}, I_p} - E_{\rm tun} S_{{\subD}, I_p} )
G^{\rm B} (E_{\rm tun})_{I_p, J_q}
( H_{J_q, {\subA}} - E_{\rm tun} S_{J_q, {\subA}} ) ,
\label{eq:TDABGF}
\end{equation}
in which the sums over $I_p$ and $J_q$ both exclude $\phi_{\subD}$ and $\phi_{\subA}$.

In the effective Hamiltonian of Eq. (\ref{eq:Heff}),
the energy $E$ is generally unknown.
However, in the application to the two-state ET case of Eq. (\ref{eq:TDABGF}),
the energy $E_{\rm tun}$ is the electron tunneling energy 
that is most naturally defined as the average of
the orbital energies of donor and acceptor orbitals,
\(
E_{\rm tun} = \left( \varepsilon_{\subD} + \varepsilon_{\subA} \right) / 2
\) .

Because our method employs the FMOs as the basis functions,
in contrast with the previous methods that employ atomic orbitals,
we can directly extract pictures that reflect the chemical properties 
of the molecular fragments.
This also applies to the tunneling current analysis to be described next.

\subsection{Electron Transfer Pathway Analysis}
\label{subsec:etpath}

There exist a number of ET pathway analysis methods.
For reviews, see Refs. \cite{SkourtisBeratan1999,ReganOnuchic1999,Stuchebrukhov2003}.
The most primitive (and thus useful in practice) would be the \textit{Pathways} model \cite{Beratan1991}
based on an empirical Green function method,
whereas one of the most sophisticated at present would be that based on
the \textit{ab initio} multi-configuration self-consistent field (MCSCF) method
with occupation restricted multiple active space (ORMAS) model \cite{NishiokaAndo2011A}.
Here we employ the tunneling current method \cite{Stuchebrukhov2003} 
originally developed with the
semi-empirical MOs
and based on atomic orbitals (AOs).
By contrast, our implementation is based on the \textit{ab initio} FMO-LCMO calculations
which provides direct picture based on the molecular fragments with reduction of the
number of basis functions compared to the AO-based methods.
In addition, it allows systematic improvements of approximations, for instance,
with exploits of recently emerging developments of density functional theories (DFTs).
This last aspect is, however, out of the scope of this article.

In the tunneling current analysis, 
the transfer integral $\TDA$ is expressed in terms of contributions
from electron current $J_{I_p, J_q}$ between basis FMOs $\{ \varphi^{I}_{p} \}$,
\begin{equation}
\TDA = \hbar
\sum_{I \in \Omega_{\subD}}
\sum_{J \notin \Omega_{\subD}}
J_{I,J} ,
\end{equation}
\begin{equation}
J_{I,J} = 
\sum_{I_p}
\sum_{J_q}
J_{I_p, J_q} ,
\end{equation}
in which the summation over $I_p$ and $J_q$ are 
over the FMOs within fragments $I$ and $J$,
and $\Omega_{\subD}$ denotes the spatial region assigned to the donor of ET.
The inter-orbital current $J_{I_p, J_q}$ is computed from
the electronic Hamiltonian and overlap matrices and
the coefficients of FMO-LCMO,
$\{ C^\supi_{I_p} \}$ and $\{ C^\supf_{I_p} \}$,
that represent the mixing of bridge FMOs to 
the donor and acceptor FMOs, $\varphi_{\subD}$ and $\varphi_{\subA}$,
in the initial (\supi) and final (\supf) states of the ET, $\psi^\supi$ and $\psi^\supf$.
They are expressed as
\begin{equation}
| \psi^\supi \rangle = 
C^\supi_{\subD} | \varphi_{\subD} \rangle +
\sum_I^N
\sum_{I_p}
C^\supi_{I_p} | \varphi^I_{p} \rangle ,
\end{equation}
\begin{equation}
| \psi^\supf \rangle = 
C^\supf_{\subA} | \varphi_{\subA} \rangle +
\sum_I^N
\sum_{I_p}
C^\supf_{I_p} | \varphi^I_{p} \rangle ,
\end{equation}
\begin{equation}
J_{I_p, J_q} = \frac{1}{\hbar}
\left( H_{I_p, J_q} - E_{\rm tun} S_{I_p, J_q} \right)
\left( C^\supi_{I_p} C^\supf_{J_q} - C^\supf_{I_p} C^\supi_{J_q} \right) .
\end{equation}
All these are thus computed straightforwardly from 
the FMO-LCMO calculation.

The tunneling current method enables us to analyze long-distance ET pathways in real-space.
As noted above, an advantage of our implementation stems from the use of FMOs as the basis functions.
Although it is possible to carry out similar analysis
with the conventional AO-based methods
by simply taking the sum within each fragments,
the advantage of the FMOs is that
the dimension of the basis functions are about one order of magnitude smaller,
and thereby the computational cost is significantly reduced.
Therefore, large ET systems that have been only tractable by semi-empirical MO calculations
are now accessible straightforwardly 
with the \textit{ab initio} MO and DFT methods.

\section{Applications to Bacterial Photosynthetic Reaction Center}
\label{sec:App}

We have applied the computational strategy 
described in the previous section to
the ETs in bacterial photosynthetic reaction center.
In this paper, we shall focus on the ET step 
between the two quinones, MQ and UQ,
via the nonheme \fedc\ ion complex,
displayed in the lower part
of Fig. 1 and in Fig. 2(b).
As presented in Fig 2(b), the \fedc\ ion is surrounded by 
five ligands from amino-acid side-chains, four histidines (His) and one glutamate (Glu).
Because the imidazole rings of the His ligands contain spatially delocalized low-energy
$\pi$-electrons, they are likely to form the bridge states to mediate the ET.

\subsection{Experimental facts}
\label{subsec:facts}

First we summarize the experimental findings.
Because the bacterial photosynthetic reaction center is a prototype system for
studying biological energy
conversions, there exist a wealth of experimental works.
The key findings to be treated in this work pertaining to the ET from MQ to UQ via
the nonheme iron complex are:
\begin{itemize}
\item
The ET time, defined by the inverse of ET rate constant $1/\kET$, 
is in the range 25 -- 36 $\mu$s for
\textit{Blastochloris viridis} \cite{LeiblBreton1991,Mathis1992}.
\vspace*{0.5em}
\item
According the electron paramagnetic resonance (EPR) experiment, 
the \fedc\ ion of the nonheme iron complex
is in the high-spin ($S=2$) state \cite{FeherOkamura1999}.
\vspace*{0.5em}
\item
Substituting the \fedc\ ion by a \zndc\ did not much alter the ET rate:
the ET time of 150 $\mu$s for the \fedc\ complex 
was reduced slightly to 140 $\mu$s for the \zndc\ complex
in \textit{Rhodobacter Sphaeroides} R-26.1 \cite{Debus1986}.

Note that this experiment was performed on a different bacteria
so that the ET time for the \fedc\ complex is different from that
in \textit{Blastochloris viridis}.
\vspace*{0.5em}
\item
Depletion of the \fedc\ ion slowed down the ET only approximately twice,
from 150 $\mu$s to 350 $\mu$s of the ET time,
in \textit{Rhodobacter Sphaeroides} R-26.1 \cite{Debus1986}.
\end{itemize}
From these findings, it has been conjectured that the \fedc\ ion
is not playing vital roles to mediate the ET from MQ to UQ.

\subsection{Computational results}

Now we present the results of computational analysis.
The focus of this paper will be on the key experimental findings listed in
Sec. \ref{subsec:facts}.
Further details have been presented in Ref. \cite{Kitoh-Nishioka2012}.
The ET pathway analysis will be demonstrated 
particularly useful for elucidating the mechanism behind
the insensitivity of ET rate on the metal ion substitution and depletion.
Before proceeding to the pathway analysis, we shall assess
the accuracy of the computed ET rate.

\subsubsection{Evaluation of electron-transfer rate}

First we
evaluate the ET rate of Eq. (\ref{eq:kET}).
As noted in Sec. \ref{sec:kET}, the formula depends on three
parameters, the reaction free energy $\Delta G$,
the nuclear reorganization energy $\lambda$,
and the electronic transfer integral $\TDA$.
While $\TDA$ is a purely electronic quantity
and $\lambda$ originates mostly from the nuclear rearrangements,
$\Delta G$ includes both the electronic redox energies 
and the nuclear relaxation energies.
Although $\lambda$ and $\Delta G$ can be evaluated
in principle with use of molecular dynamics simulations,
their accuracy depends crucially on the reliability of 
the force-field model for the entire protein system,
which has not been well established.
Therefore, we employ experimentally evaluated values of
$\lambda = 1.0$ eV and $\Delta G = -0.07$ eV for a particular bacteria
\textit{Blastochloris viridis} \cite{Moser1992,Moser1995},
and place our focus on $\TDA$ which is the least straightforward for
experimental evaluations.

The nuclear coordinates were taken from the X-ray crystallographic data
in the Protein Data Bank (code 1PRC) \cite{Deisenhofer1995}.
Those of hydrogen atoms were optimized
with the semi-empirical PM3 quantum chemical calculations,
in which the heavier atoms were fixed at the crystallographic coordinates.
For the wild-type (WT) complex with the high-spin ($S=2$) \fedc\ ion,
the unrestricted Hartree-Fock (UHF) method was used.
For other systems in the low-spin ($S=0$) state,
the restricted HF (RHF) method was used.
In addition, for these low-spin cases,
calculations with the FMO-LCMO method described in Sec. \ref{subsec:fmolcmo}
were carried out
in an aim to assess its accuracy.
Moreover, we also assessed the FMO-LCVMO method \cite{NishiokaAndo2011B} that limits
the monomer FMO space to the minimal valence (VMO) space.
The 6-31G(d) basis set was employed throughout.

The computed results of the transfer integral $\TDA$ 
and the ET time $1/\kET^{\rm calc}$
are listed in Table \ref{tbl:1}.
For the WT system with high-spin \fedc,
the computed $\TDA$ with the GMH and the BGF methods were
1.33 \wn\ and 1.39 \wn, respectively.
The agreement between the two methods supports their accuracy and consistency.
It demonstrates in particular that the BGF method with the
appropriate tunneling energy $E_{\rm tun}$ is capable of capturing the $\TDA$ value
corresponding to the diabatic surface crossing,
similarly to the GMH method that was thus designed.

With use of these $\TDA$ values,
together with $\Delta G$ and $\lambda$ from the experimental
evaluation as described above,
the ET time $1/\kET^{\rm calc}$ was
calculated to be 8.7 -- 9.5 $\mu$s at the room temperature.
As noted in Sec. \ref{subsec:facts},
the observed ET time
was in the range 25 -- 36 $\mu$s.
Thus, the computed values
a few times underestimate the experimental.
However, the agreement of this order is sufficiently reasonable 
since the rate constant is exponentially sensitive to
the thermal activation factor.
Indeed, 
higher accuracy cannot be expected 
even with the state-of-the-art quantum chemical computation.
In this regard, it is important not to rely too much on a single number
but to carry out analysis from multiple viewpoints.
To this end, we next discuss analysis on
the spin state alteration and metal ion substitution.

\begin{table}
\caption{Computed electron transfer integral $|\TDA|$ (in \wn) 
and electron transfer time $1/\kET^{\rm calc}$ (in $\mu$s) 
for \textit{Blastochloris viridis}
with various metal ions and various computational methods.
For the observed electron transfer time $1/\kET^{\rm obs}$,
the values in the parentheses are for another bacteria 
\textit{Rhodobacter Sphaeroides} R-26.1.
See the text for details.
}
\label{tbl:1}
\begin{tabular}{p{2cm}p{2cm}p{1.5cm}p{1.5cm}p{1.5cm}p{1.5cm}}
\hline\noalign{\smallskip}
metal ion & method & \hspace*{3em} & $|\TDA|$ & $1/\kET^{\rm calc}$ & $1/\kET^{\rm obs}$ \\
\noalign{\smallskip}\hline\noalign{\smallskip}
\fedc\      & UHF       & GMH     & 1.33  & 9.48 & 25--36$^a$ \\
(high-spin) &           & BGF     & 1.39  & 8.68 & (150)$^b$ \\
\noalign{\smallskip}\hline\noalign{\smallskip}
\fedc\      & RHF       & GMH     & 0.944 & 18.7 & \\
(low-spin)  & FMO-LCMO  & GMH     & 0.955 & 18.3 & \\
            &           & BGF     & 0.979 & 17.4 & \\
            & FMO-LCVMO & GMH     & 0.766 & 28.4 & \\
            &           & BGF     & 0.777 & 27.6 & \\
\noalign{\smallskip}\hline\noalign{\smallskip}
\zndc\      & RHF       & GMH     & 1.30  & 9.93 & (140)$^b$ \\
            & FMO-LCMO  & GMH     & 1.31  & 9.65 & \\
            &           & BGF     & 1.34  & 9.21 & \\
            & FMO-LCVMO & GMH     & 1.20  & 11.5 & \\
            &           & BGF     & 1.23  & 11.1 & \\
\noalign{\smallskip}\hline\noalign{\smallskip}
none        & RHF       & GMH     & 0.610 & 44.8 & (350)$^b$ \\
            & FMO-LCMO  & GMH     & 0.649 & 39.4 & \\
            &           & BGF     & 0.655 & 38.8 & \\
            & FMO-LCVMO & GMH     & 0.731 & 31.2 & \\
            &           & BGF     & 0.737 & 30.7 & \\
\noalign{\smallskip}\hline\noalign{\smallskip}
\end{tabular}
\\
$^a$ For \textit{Blastochloris viridis} \cite{LeiblBreton1991,Mathis1992}.
\\
$^b$ For \textit{Rhodobacter Sphaeroides} R-26.1 \cite{Debus1986}.
\end{table}

\subsubsection{Role of nonheme iron complex: spin-state and substitution}

While the EPR experiment indicated that the WT complex is in the
high-spin ($S=2$) state, it would be still intriguing to examine different spin states
in an aim to explore their functional significance.
This is straightforward for the computational analysis.
In general, the metal-ligand distances in \fedc\ complexes can vary
by a few tenth of \AA\ 
in different spin states.
However, we used here the same molecular structure as that of the high-spin state
in order to focus on the effect of spin state 
without introducing extra factors.

In the low-spin ($S=0$) state,
the transfer integral $\TDA$ was calculated to be in a range 0.94 -- 0.98 \wn\ 
with the full RHF and the FMO-LCMO methods.
On the technical aspect, we note the accuracy of the FMO-LCMO methods in comparison with
the full RHF reference.
The smaller $\TDA$ of the low-spin state than that of the high-spin state
results in approximately twice slower ET rate
with the ET time of 17 -- 19 $\mu$s.
It is unclear at present if 
the faster ET rate in the WT high-spin state has some physiological significance.
The difference of the factor 2 is rather modest.
Nevertheless, because the \fedc\ complex can change the spin state depending
on the ligand structure, this small difference may
have a functional significance to feature a possibility to control the ET rate.
This aspect would be open for further investigations.
Finally, we note that the FMO-LCVMO method, 
limited to the minimal valence MO space, 
gives reasonable values of $\TDA = $ 
0.77 -- 0.78 \wn.

Next we substituted the \fedc\ ion by a \zndc\ ion.
For the same reason as that noted above for the spin-state variation,
we used the same molecular structure as that of the high-spin WT state.
The computed transfer integral $\TDA$ was 1.30 -- 1.34 \wn\
with the full RHF and the FMO-LCMO methods.
These are very close to the values for the high-spin WT case.
As a result, the computed ET times for the \zndc\ complex, 9.2 -- 9.9 $\mu$s,
were also very close to those for the high-spin WT case.
This is in good accord with the experimental findings
for \textit{Rhodobacter Sphaeroides} R-26.1:
the ET time of 150 $\mu$s for the WT system compared to 
140 $\mu$s for the \zndc-substituted system. 
Thus, the insensitivity of the ET rate against the metal ion substitution
is well reproduced.
[Because the experimental ET rate of \zndc-substituted system
was unavailable
for \textit{Blastochloris viridis},
we compared the ratio of the ET times 
for \textit{Rhodobacter Sphaeroides} R-26.1.
This also applies to the metal-depleted case discussed next.]

Finally, we carried out computation without the metal ion.
Again, we used the same molecular structure as that of the high-spin WT state
in an aim to extract the essential roles of the \fedc\ ion,
even though 
the molecular structure must have reorganized
in the actual metal depleted system in the experiment.
The computed transfer integral $\TDA$ was 0.61 -- 0.66 \wn,
about twice smaller than that for the WT high-spin \fedc\ system.
Consequently, the computed ET time was 39 -- 45 $\mu$s, about four times
slower than the WT system. 
This is again in qualitative accord with the experimental finding
for \textit{Rhodobacter Sphaeroides} R-26.1:
the ET time of 350 $\mu$s for the metal depleted system compared to 150 $\mu$s for the WT system.

\begin{figure}[b]
\begin{center}
\includegraphics[width=0.6\textwidth]{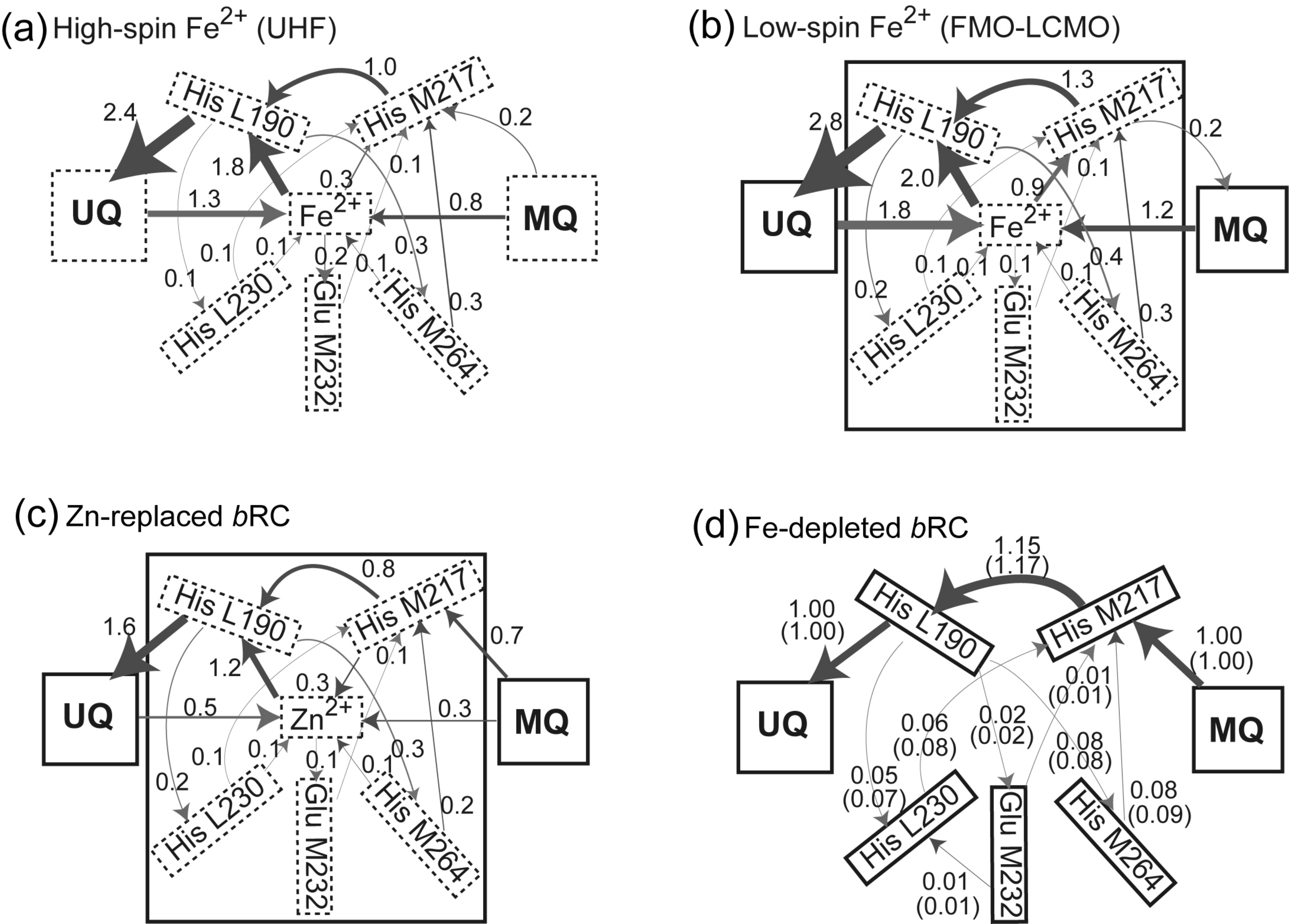}
\caption{Results of electron transfer pathway analysis for
(a) wild-type high-spin ($S=2$) \fedc\ complex,
(b) low-spin ($S=0$) \fedc\ complex,
(c) the ion complex with the \fedc\ ion replaced by a \zndc\ ion,
and
(d) the complex without the metal ion.
The numbers near the arrows denote the normalized ET currents
$\hbar J_{I,J} / \TDA$
between the fragments $I$ and $J$,
which are also approximately represented by the thickness of the arrows.
}
\label{fig:3}
\end{center}
\end{figure}

\subsubsection{Electron transfer pathway analysis}

To gain further insights into the mechanism of ET,
particularly with respect to the role of the nonheme iron complex,
we carried out ET pathway analysis described in Sec. \ref{subsec:etpath}.
Figure 3 displays the computed tunneling currents, in which 
the width of the arrows represents the relative contribution of the path
measured by the normalized tunneling current 
$\hbar J_{I,J} / \TDA$.
As noted in Sec. \ref{subsec:facts}, 
the experimentally found insensitivity of the ET rate 
on the substitution and depletion of the \fedc\ ion have raised a conjecture
that the \fedc\ ion does not contribute to the major ET pathway.
However, the computational result in Fig. 3(a) indicates the contrary: 
the major ET pathway consists of MQ $\to$ \fedc $\to$ His L190 $\to$ UQ.
The resultant patterns of pathways for the high-spin (a) and low-spin (b) cases
are basically very similar, but
the widths of the main arrows are overall wider in the low-spin (b) case.
Nonetheless, because the contribution of the back flow from UQ to \fedc\ 
is also larger in (b) than in (a),
the net ET current is smaller in the low-spin system.

In the \zndc-substituted system displayed in Fig. 3(c), 
both the forward current from MQ to \zndc\ and the backward current from UQ to \zndc\
are reduced compared to the \fedc\ cases in Fig. 3(a) and (b).
This implies that the filled (3d)$^{10}$ AO configuration of \zndc\ is less 
effective to mediate the ET than the unfilled (3d)$^{6}$ of \fedc.
However,
the contribution of a pathway from MQ to His M217 becomes larger in the \zndc\ system,
as if to compensate the reduced current through the metal ion.
In addition, the direction of the current from the metal ion to His M217
is reversed such that the current is now directed from His M217 to \zndc.
These elucidate the electronic mechanism behind the insensitivity
of the transfer integral $\TDA$ and the ET rate on
the substitution of \fedc\ by \zndc.

Figure 3(d) displays the corresponding pathway analysis for
the metal ion depleted system.
Now it is clear that the pathway via His M217 and His L190 becomes dominant.
Consequently, the reduction of the ET rate is only by a factor of 3 -- 5,
in accord with the experimental observation.

In this way, an intriguing picture emerges such that the His ligands play the role to secure
the robustness of ET by providing an auxiliary ET pathway channel even 
in cases of
disturbances such as the metal ion substitution or depletion.

\section{Concluding Remarks}
\label{sec:concl}

A new computational scheme to carry out ET pathway analysis in large biomolecules
has been developed
and applied to the ET from MQ to UQ via a nonheme iron ion complex
in bacterial photosynthetic reaction center.
The scheme consists of a combination of the FMO-LCMO method that enables
\textit{ab initio} electronic structure calculations of large biomolecules
and the tunneling current analysis that provides pictorial understanding
of ET mechanism.
Nevertheless, since any computational studies on realistic molecular systems generally involve
assortments of theoretical and computational approximations, 
it is essential to proceed with sufficient care checking the consistency
with experimental findings.
Then, it will become possible to discuss detailed microscopic mechanism at
atomic and electronic levels that are inaccessible by experiments.
With such intimate collaborations of experimental and theoretical studies,
the progress of our understanding of biological energy conversions
will be secured.

\bibliographystyle{spphys}

\end{document}